# On the Dynamics of Generalized Coherent States. I.
## Exact and Stable Evolution

B.A. Nikolov, D.A. Trifonov

**Abstract.** The exact and stable evolutions of generalized coherent states (GCS) for quantum systems are considered by making use of the time-dependent integrals of motion method and of the Klauder approach to the relationship between quantum and classical mechanics. It is shown that one can construct for any quantum system overcomplete family of states (OFS), related to the unitary representations of the Lie group $G$ by means of integral of motion generators, and the possibility of using this group as a dynamical symmetry group is pointed out. The relation of the OFS with quantum measurement theory is also established.

## 1 Introduction

The generalized coherent states (GCS) were introduced in principle by J. Klauder [1] as an example of his overcomplete family of states (OFS), which realize continuous representations of Hilbert space $\boldsymbol{H}$ and are convenient for the study of the general relationship between classical and quantum mechanics. In the second paper [1] an OFS was defined as

$$|\Phi(\ell_a)\rangle = V(\ell_a)|\Phi_0\rangle = \exp(\ell_a L_a)|\Phi_0\rangle \qquad (1)$$

where $|\Phi_0\rangle$ is a fixed ("fiducial") vector in $\boldsymbol{H}$ and $V(\ell_a)$ is a unitary irreducible representation (UIR) in $\boldsymbol{H}$ of a Lie group $G$, $\ell_a$ being the canonical group parameters; and $L_a$, the generators of the representation $V(\ell_a)$, $a = 1, 2, \ldots, r$. In Ref. 1 it was noted that states (1) form automatically an OFS for UIR of compact groups, but the example that has been thoroughly examined is connected with the representation operator

$$V(q, p) = \exp(-iqP + ipQ), \quad [Q, P] = i, \qquad (2)$$

of the Weyl-Heisenberg group $G_W$, and

$$|\Phi(q, p) = V(q, p)|\Phi_0\rangle. \qquad (3)$$

The same OFS was independently studied by Glauber [2] in the notations

$$|\alpha\rangle = D(\alpha)|\Phi_0\rangle, \quad D(\alpha) = \exp(\alpha a^\dagger - \alpha^* a) = V(q, p), \qquad (4)$$

where the parameters $q$ and $p$ are, to within a factor, the real and imaginary parts of the complex number $\alpha$, and $a$, $a^\dagger$ are the known lowering and raising operators for the harmonic oscillator [1]. Glauber has called such states *coherent* (CS). The CS (3), (4) (further called *usual CS* or $G_W$-CS) possess a number of remarkable physical and mathematical properties and later many authors [3-10] (see also [11-13] and references therein) have introduced and studied more general CS, generalizing different properties of the usual CS but one of the best generalizations is just the first one, proposed by Klauder by Eq. (1). The (over) completeness of the vectors (1), as it was shown in Ref. 7, holds for any UIR of a Lie group $G$. In the same paper [7] a factorization of the OFS (1) over the stationary group $K \subset G$ of the fixed vector $|\Phi_0\rangle$ was also proposed since the physical states are defined up to a constant phase factor:

$$V_h|\Phi_0\rangle = e^{i\alpha(h)}|\Phi_0\rangle, \quad h \in K \subset G, \tag{5}$$

$$|\Phi_x\rangle = V_{s(x)}|\Phi_0\rangle, \quad x \in X = G/K, \tag{6}$$

where $s$ is a cross-section in the group fibre bundle $(G, X, \pi)$ [14], i.e., $s(x)$ is a representative of the coset, $x = gK$, $g \in G$. Such factorized OFS have been called system of GCS. We shall use in this paper both terms.

The properties and applications of CS and GCS are widely discussed in the literature [1-13] so we would not list them here. We shall only elucidate in Sec. 2 the geometrical meaning of GCS as images of cross-sections in fibre bundle $(M, X, \rho)$ associated with the fibre bundle $(G, X, \pi)$. The relation with the X-measurements in sense of Holevo [15] is pointed out also.

In our paper we discuss two main aspects of the dynamics of GCS – their exact time evolution and their stable evolution. By stable evolution we mean that the initial GCS, when evolves in time , remains a GCS of the same type at all times, i.e., the set of GCS is invariant under time evolution.

The dynamics of the usual CS is completely examined [1,5,11,16-21]. Their exact evolution was explicitly found for quadratic systems [5,20,21] and in Ref. 22 a method was proposed to construct usual CS for any system. The stable evolution of CS was considered by several authors [16-19]. The most general form of the Hamiltonian which preserves all CS stable was found at first in Refs. 16,17 and later by other methods in Refs. 18,19.

The main results of the present paper are the following. In Sec. 3 we give a method for constructing GCS for any Lie group and any quantum system by the realization of representation $V_g$ of $G$ in the space $\boldsymbol{H}$ of solutions of the Schrödinger equation. For this purpose one may use invariant lowering and raising operators $A(t)$, $A^\dagger(t)$ [5,11] and express $V_g$ in terms of them [23-25]. This method permits one also to conclude that the dynamical symmetry group of quantum system may be any group $G$, which has UIR in the Hilbert space $\boldsymbol{H}$. Using the sufficient conditions (21), (22) we prove that the $N$-dimensional oscillator preserves stable CS for any Lie group $G$, the representations of which are generated by the operators $a_i a_j$, $a_i^\dagger a_j^\dagger$ ($i, j = 1, 2, \ldots, N$). Some applications of the developed method are considered in the subsequent paper (see Ref. 26).

---

[1] and $|\Phi_0\rangle$ being its ground state $|0\rangle$: $a|0\rangle = 0$. (Note, added in the e-print quant-ph/0407260).

## 2 GCS and Quantum Measurements

In this section we briefly consider the factorization [7] of the OFS (1) and elucidate the geometrical meaning of GCS defined by Eq. (6) as well as the relation of the systems GCS with quantum measurement theory [15,27].

Let $K$ be stationary subgroup of the vector $|\Phi_0\rangle$, $V_h|\Phi_0\rangle = e^{i\alpha(t)}|\Phi_0\rangle$, $h \in K \subset G$. Then it is clear that two vectors $|\Phi_{g_1}\rangle$ and $|\Phi_{g_2}\rangle$ differ from each other by a phase factor iff $g_2 = g_1 h$, i.e., iff $g_1$ and $g_2$ belong to the same coset $gK$. The state is determined by the point $x \in X = G/K$. One can construct an OFS according to (1) choosing a representative $s(x)$ from every coset $gK$. The so-obtained OFS factorized in this manner was called a system of GCS [7]. The function $s(x)$ is evidently a cross-section in the group fibre bundle $(G, X, \pi)$, $\pi$ being the canonical projection. For a given $s(x)$ every element $g \in G$ can be written as a product $g = s(x_g)h_g$, where $x_g$ denotes the coset to which $g$ belongs. Thus

$$|\Phi_g\rangle = e^{i\alpha(h_g)}|\Phi_{x_g}\rangle. \tag{7}$$

The operators of the representation $V_g$ act on $|\Phi_x\rangle$ transitively to within a phase factor:

$$V_g|\Phi_x\rangle = V_g V_{s(x)}|\Phi_0\rangle = \exp(if(g,x))|\Phi_{x'}\rangle, \tag{8}$$

where $x'$ is determined from the equation $gs(x') = s(x')h(g,h)$. Introducing the mappings $\delta: G \times K \to G$, $\delta(g,h) \equiv \delta_h(g) = gh$ and $\sigma: \mathcal{H} \times K \to \mathcal{H}$, $\sigma(|\psi\rangle, h) \equiv \sigma_h(|\psi\rangle) = e^{i\alpha(h)}|\psi\rangle$, where $\alpha(h)$ is defined by Eq. (4) ($\mathcal{H}$ being the set of unit vectors in $\boldsymbol{H}$) one can consider $G$ and $\mathcal{H}$ as $K$-manifolds. The following diagram

$$\begin{array}{ccc} G \times K & \xrightarrow{\delta} & G \\ \Phi \times id \downarrow & & \downarrow \Phi \\ \mathcal{H} \times K & \xrightarrow{\sigma} & \mathcal{H} \end{array} \quad , \quad \Phi: g \to |\Phi_g\rangle \tag{9}$$

is commutative and consequently the map $\Phi: G \to \mathcal{H}$ is a morphism in the category of $K$-manifolds [14]. The set of such morphisms $\mathrm{Hom}_K(G, \mathcal{H})$ is in one-to-one correspondence with the set of cross-sections $\tilde{\Phi}$ in the associated fibre bundle $(M, X, \rho)$, where $M = (\mathcal{H} \times G)/K$ and $\rho: M \to X$ is the canonical projection: $\rho \circ \gamma = \pi \circ \mathrm{pr}_1$ [14]. Here $\gamma := G \times \mathcal{H} \to M$ satisfy the following condition: $\gamma(g, |\psi\rangle) = \gamma(gh, \sigma_h(|\psi\rangle))$, $h \in K$. The correspondence $\Phi \to \tilde{\Phi}$ is provided by the commutative diagram

$$\begin{array}{ccc} G & \xrightarrow{id \times \Phi} & G \times \mathcal{H} \\ \pi \downarrow & & \downarrow \gamma \\ X & \xrightarrow{\tilde{\Phi}} & M \end{array} \tag{10}$$

From this diagram it is easy to obtain

$$\tilde{\Phi}(x) = \gamma(g, |\Phi\rangle), \quad g \in \pi^{-1}(x). \tag{11}$$

Defining $\gamma(g, |\psi\rangle) = \exp(-i\alpha(h_g))|\psi\rangle$ we may identify the system of GCS $|\Phi\rangle$ with the image of the cross-section $\tilde{\Phi}$.

The most important property of the system GCS is their overcompleteness, expressed by the equation

$$\int \mu(dx)|\Phi_x\rangle\langle\Phi_x| = 1, \qquad (12)$$

where $\mu(dx)$ is an invariant measure on $X$. Eq. (12) implies that the family of operators

$$M(\Delta) = \int_\Delta \mu(dx)|\Phi_x\rangle\langle\Phi_x|, \quad \Delta \in \mathcal{B}(X),$$

($\Delta$ being a Borel set in $X$, $\mathcal{B}$ - $\sigma$-algebra of Borel sets) form a generalized resolution of identity (or *positive operator-valued measure* on $X$) [15]. According to a theorem due to Davies and Lewis [27] there exist at least one measurement $\mathcal{E}$, determined by the formula

$$\mathrm{tr}(\mathcal{E}_\Delta(\rho)) = \mathrm{tr}(\rho M(\Delta)). \qquad (13)$$

Recall that the measurement $\mathcal{E}$ on $X$ is a linear continuous map $\mathcal{E} : \mathcal{B}(X) \to \mathrm{Aut}(\mathcal{T}(\boldsymbol{H}))$, $\mathcal{T}(\boldsymbol{H})$ being the space of trace class operators in $\boldsymbol{H}$ such that $\mathrm{tr}\mathcal{E}_X(\rho) = \mathrm{tr}\rho$ and the positive cone $\mathcal{T}(\boldsymbol{H})^+$ is invariant under $\mathcal{E}_\Delta$, $\Delta \in \mathcal{B}(X)$ [28]. Basing on Eq. (13) the (generalized) resolution of identity $M(\Delta)$ itself can be referred to as $X$-measurement [15] (Let us note that the Holevo definition differs from the one given above). Moreover it is clear from the construction that this measurement is *covariant* with respect to UIR $V_g$, i.e. the following condition is fulfilled [15]

$$V_g^* M(\Delta) V_g = M(g^{-1}\Delta), \quad \Delta \in \mathcal{B}(X). \qquad (14)$$

If the state of a quantum system is described by the density operator $\rho$, then the probability distribution of the results of the measurement $M(\Delta)$ is given by Eq. (13). The covariance property (14) permits one to establish a relation between the physical characteristics of the system and the resolutions of identity in Hilbert space $\boldsymbol{H}$.

## 3 Evolution of GCS

The dynamics of quantum states is determined by the evolution operator $S_t$:

$$S_t = \mathrm{T}\exp\left(-\int_0^t H(t)dt\right),$$

where $H(t)$ is the Hamiltonian of the system. The evolution of an OFS is given by the relation $|\Phi_g; t\rangle = S_t|\Phi_g\rangle$. This direct action by $S_t$ on the vectors $|\Phi_g\rangle$ may turn to be difficult even impossible especially in the case of nonstationary Hamiltonians. It is proved to be more effective the method of integrals of motion, developed in Refs. 5,11.

If $A$ is an integral of motion then it commutes with the Schrödinger operator $D_S = i\partial_t - H$, $[A, D_S] = 0$, and thereby one can get new solutions acting by $A$ on a fixed solution $|\Phi_0; t\rangle$. Thus if the operators of the representation $V(\ell_a) = V_g(t)$ are realized as integrals of motion, then by means of them one can obtain OFS (1) as solutions of the equation of motion according to the formulae

$$|\Phi_g(t)\rangle = V_g(t)|\Phi_0(t)\rangle = \exp(\ell_a L_a(t))|\Phi_0(t)\rangle$$
$$= S_t V_g S_t g^{-1}|\Phi_0\rangle = S_t|\Phi_g\rangle = |\Phi_g; t\rangle. \qquad (15)$$

Here the generators $L_a(t) = S_t L_a S_t^{-1}$ are formal solutions of the equation $[A, D_S] = 0$ and consequently are integrals of motion.

Because of the completeness of system of vectors (15) every solution of the Schrödinger equation may be realized in the carrier space $\boldsymbol{H}$ of UIR $V_g(t)$. The group with such a property is called *dynamical symmetry group* of the quantum system. For nonstationary systems the dynamical symmetry was studied in Refs. 11,29.

Thus if the vectors $|\Phi_g\rangle = V_g|\Phi_0\rangle$ form an OFS in Hilbert space $\boldsymbol{H}$ then the related group $G$ may serve as a group of dynamical symmetry for any quantum system. This assertion is in agreement with the results of Refs. 11,29, where it was shown that for $N$-dimensional system the noncompact group $U(N, 1)$ ca describe dynamical symmetry. Now we get from Eq. (15) that the dynamical symmetry can be described by any Lie group $G$, which has UIR in $\boldsymbol{H}$.

In practical calculations the more efficient way is to solve equation $[A, D_S] = 0$, looking for solutions $A(t)$ of some special form, say linear in generators $L_a$. Since $L_a$ may be expressed in terms of lowering and raising boson operators $a$, $a^\dagger$, it is convenient to construct first the invariant operators

$$A = S_t a S_t^{-1}, \quad A^\dagger = S_t a^\dagger S_t^{-1} \qquad (16)$$

and then to use them for construction of generators $L_a(t)$. We follow this way in subsequent paper [26] in constructing exact time evolution of OFS for some groups and quantum systems. Here we would like to say that integrals of the form (16) are explicitly constructed for any quadratic Hamiltonians [20,21] and for some nonquadratic ones [11].

Let us turn to the question of stable evolution of OFS. The Klauder condition $S_t \mathcal{C} \subset \mathcal{C}$ ($\mathcal{C}$ being the manifold of GCS) may be written more explicitly in the form:

$$S_t V(\ell_a)|\Phi_0\rangle = S_t|\Phi(\ell_a)\rangle = |\Phi(\ell_a(t))\rangle = V(\ell_a(t))|\Phi_0\rangle, \qquad (17)$$

i.e., the whole time-dependence of vectors from OFS is contained in the group parameters $\ell_a(a)$. It is apparent that Eq. (17) holds for any fixed vector $|\Phi_0\rangle$ iff $S_t$ is an operator of the same representation: $S_t = V_{g(t)}$. Moreover, I.A. Malkin has proved the following

*Theorem* [25]. Arbitrary system of GCS (1) remains stable under time evolution iff the Hamiltonian of the quantum system has the form

$$H = f_a(t) L_a, \qquad (18)$$

where $f_a(t)$ are arbitrary functions.

On the face of it this theorem contradicts to the well known fact [16-19] that the most general form of the Hamiltonian which preserves stable the system of usual CS is

$$H = \omega(t) a^\dagger a + F(t) a^\dagger + F^*(t) a + \beta(t), \qquad (19)$$

which is nonlinear in $a$, $a^\dagger$. Let us note however that the operators $a^\dagger a$, $a$, $a^\dagger$, $1$ form a projective representation of Lie algebra of the two-dimensional Euclidean group $E(2) = T^2 \otimes SO(2)$. Then using the relation

$$\exp\left(\lambda\, a^\dagger a + \mu a + \nu\, a^\dagger\right) = \exp(u\, a^\dagger)\exp(z\, a^\dagger a)\exp(va)\exp(w),$$

where $z$, $u$, $v$, $w$ depend on the canonical parameters $\lambda$, $\mu$, $\nu$, one can see that GCS for this representation of $E(2)$ and $|\Phi_0\rangle = |0\rangle$ coincide with the usual CS. The Hamiltonian (18), predicted by Malkin theorem for $E(2)$, has just the form (19).

Let us now consider the special case when the fiducial vector $|\Phi_0\rangle$ is stable under the action of the evolution operator $S_t$:

$$S_t|\Phi_0\rangle = |\Phi_0\rangle. \tag{20}$$

Then for the stability of OFS $(V_g, |\Phi_0\rangle)$ the evolution operator $S_t$ ought not be operator from the representation $V_g$. $S_t$ can be an (external) automorphism of the group of representation operators:

$$S_t V_g S_t^{-1} = V_{g(t)}, \quad g = (\ell_a). \tag{21}$$

The OFS $(V_g, |\Psi_0\rangle)$ for which the conditions (20), (21) are satisfied may be called *superstable* relatively to $S_t$ (or to corresponding Hamiltonian). Another explanation of the above-mentioned seeming contradiction can be given if one notes that the system of usual CS is superstable relatively to Hamiltonian (19).

A superstable OFS may be realized also when the generators of the representation are homogeneous functions of $a$, $a^\dagger$ and $S_t$ is the evolution operators corresponding to the harmonic oscillator. Then

$$\exp(\ell_a L_a(t)) = \exp(\ell_a(t) L_a) \tag{22}$$

which can be easily proved using the formula [30]

$$\exp(s\, a^\dagger a) F(a, a^\dagger) \exp(-s a^\dagger a) = F\left(ae^{-s}, a^\dagger e^s\right). \tag{23}$$

Obviously this result can be easily extended to the case of $N$-dimensional oscillator. If $X_i$ ($i = 1, 2, \ldots, p+q$) are generators of the lowest $(p+q)$-representation of Lie group $U(p,q)$, then the following Hermitian[2] operators

$$\begin{aligned}
L_i = \tilde{\phi} X_i \phi, \qquad &\phi = (a_1, \ldots, a_p, a_{p+1}^\dagger, \ldots, a_{p+q}^\dagger)^{\mathrm{T}}, \\
&\tilde{\phi} = (a_1^\dagger, \ldots, a_p^\dagger, -a_{p+1}, \ldots, -a_{p+q})
\end{aligned} \tag{24}$$

are homogeneous generators of $U(p,q)$.[3] Thus we derive that $N$-dimensional oscillator preserves stable GCS $(V_g, |\Phi_n\rangle)$ for any Lie group $G$, the representation $V_g$ being generated by the corresponding subset of operators (24), and $|\Phi_n\rangle$ - any stationary state.

The stable evolution of OFS is correctly determined by the functions $\ell_a(t)$, which are solutions of Euler equations for the functional [1]

$$I(f) = \int dt \left(i\langle f|d/dt|f\rangle - \langle f|H|f\rangle\right) \tag{25}$$

---

[2] "Hermitean" and "Weil" changed to "Hermitian" and "Weyl". (Note added).

[3] This construction by means of $N = p + q$ operators $a_i^\dagger$, $a_i$ was noted by I.T. Todorov, In: *Fizika vysokih energiy i teoria elementarnyh chastits* (Kiev, 1967) (Note added).

whose domain is restricted to the OFS-manifold, i.e., $|f\rangle = |\Phi_g\rangle$, $g = (\ell_a(t))$. Minimizing functional (25) one gets the following equations [1]:

$$\begin{aligned}
(\partial_a R_b - \partial_b R_a)\dot{\ell}_b &= \partial_a \mathcal{H}, & \partial_a &= \partial/\partial \ell_a, \\
\mathcal{H} &= \langle \Phi_g | H | \Phi_g \rangle = \mathcal{H}(\ell_a), & g &= (\ell_a), \ a = 1, 2, \ldots, r, \\
R &= (1 - \exp(-\ell_a C_a))(\ell_a C_a)^{-1} v, & v &= (v_a), \\
v_a &= i\langle \Phi_0 | L_a | \Phi_0 \rangle, & (C_a)_{bd} &= C_{ab}^d,
\end{aligned} \quad (26)$$

where $C_{ab}^d$ are structural constants of the group $G$. The classical action functional assumes the form:

$$I = \int (R_a \ell_a - \mathcal{H}). \quad (27)$$

If the matrix $\Omega_{ab} = \partial_a R_b - \partial_b R_a$ is symplectic, i.e., the 2-form $\Omega_2 = \Omega_{ab} d\ell_a \wedge d\ell_b$ is nondegenerate, closed ($d\Omega_2 = 0$) and exact ($\Omega_2 = d\Omega_1$), then the classical system described by Eqs. (26) was studied by R.M. Santilli [31] and called by him Birkhoffian system, $\mathcal{H}$ being the Birkhoffian of the system. It is not difficult to see that the matrix $\Omega_{ab}$ is symplectic iff it is nondegenerate. The group manifold in this case has symplectic structure and may be regarded as a classical phase space. Involving the matrix $\Omega^{ab}$, the inverse to the matrix $\Omega_{ab}$, one can write Eqs. (26) in the form

$$\dot{\ell}_a = (\ell_a, \mathcal{H}), \quad a = 1, 2, \ldots, r.$$

where the brackets (,) are defined by

$$(A, B) = \Omega^{ab} \partial_a A \partial_b B$$

and apparently are generalization of usual Poisson brackets [31,32].

In the case when $\Omega_{ab}$ is singular (e.g., for groups with odd dimensions [1]) the equations of motion (26) do not determine the solutions $\ell_a(t)$ uniquely. Then the dynamics of stable OFS can be effectively described by classical equations of motion in the quotient space $X = G/K$ ($K$ being the stability subgroup of $|\Phi_0\rangle$), which can be treated as phase space. The symplectic structure on $X$ was constructed [4] by E. Onofri [33]:

$$\omega = i\left(\partial^2 f / \partial z_i \partial z_j^*\right) dz_i \wedge dz_j^*, \quad (28)$$

where $z_i$ are (complex) local coordinates on $X$ and $f = f(z, z^*) = \ln |\langle \Phi_0 | V_g | \Phi_0 \rangle|^{-2}$ is the so called Kähler potential. Then on $X$ there exists a Poisson bracket

$$\begin{aligned}
(A, B) &= g^{ij} \left(\partial_i A \partial_j^* B - \partial_i^* A \partial_j B\right), \\
\partial_i A &= \partial A / \partial z_i, \quad \partial_i^* A = \partial A / \partial z_i^*,
\end{aligned} \quad (29)$$

where $g^{ij}$ is the matrix inverse to the matrix $\|i\partial^2 f / \partial z_i \partial z_j^*\|$. Consequently the equation of motion for $z_i = z_i(t)$ has the form

$$\dot{z}_i = (z_i, \mathcal{H}) = g^{ij} \partial \mathcal{H} / \partial z_j^*,$$

---

[4] for semisimple compact Lie groups. Eq. (28) holds for some noncompact groups as well, in particular for $G_W$ and $SU(1,1)$, treated in the subsequent paper [26]. (Note added in the e-print).

$\mathcal{H}$ being the classical Hamiltonian (26).

Finally we shall consider the time evolution of the probability distribution $w_\rho(\Delta) = \text{tr}(\rho M(\Delta))$ when the density operator $\rho$ evolves in time: $\rho(t) = S_t \rho S_t^{-1}$. Suppose that the system of GCS that determines the (generalized) resolution of identity $M(\Delta)$ (Sec. 2), is stable under evolution operator $S_t$. Then $S_t = V_{g(t)}$ and making use of covariance property (14) one immediately derives

$$w_{\rho(t)}(\Delta) = \text{tr}\left(\rho M(g^{-1}(t)\Delta\right) = w_\rho\left(g^{-1}(t)\Delta\right),$$

i.e. the probability distribution is only translated by means of the group transformation $g^{-1}(t)$, providing some motion in phase space $X$. Thus in this case too the quantum evolution is represented as a classical motion on the manifold $X$.

---

[5] e-print quant-ph/0407261. (Note added).

Николов Б.А., Трифонов Д.А.                    E2-81-797
**О динамике обобщенных когерентных состояний.**
**I. Точная и стабильная эволюция**

С помощью метода интегралов движения и подхода Клаудера к связи квантовой и классической механик рассмотрена точная и стабильная эволюция обобщенных когерентных состояний квантовых систем. Показано, что переполненные системы состояний квантовых систем можно строить путем реализации генераторов представлений группы Ли как интегралов движени квантовой системы. Соответствующую группу Ли, связанную с переполненной системой состояний, можно рассматривать как группу динамической симметрии физической системы. Рассмотрена связь обобщенных когерентных состояний с квантовой теорией измерений и геометрический смысл этих состояний.

Работа выполнена в Лаборатории теоретической физики ОИЯИ

Сообщение Объединенного института ядерных исследований. Дубна 1981.

Nikolov B.A., Trifonov D.A.                    E2-81-797
**On the Dynamics of Generalized Coherent States.**
**I. Exact and Stable Evolution**

The exact and stable evolutions of generalized coherent states (GCS) for quantum systems are considered by making use of the time-dependent integrals of motion method and of the Klauder approach to the relationship between quantum and classical mechanics. It is shown that one can construct for any quantum system overcomplete family of states (OFS), related to the unitary representations of the Lie group $G$ by means of integral of motion generators, and the possibility of using this group as a dynamical symmetry group is pointed out. The relation of the OFS with quantum measurement theory is also established.

Communication of the Joint Institute for Nuclear Research, Dubna 1981.